\documentclass[aps,manuscript,showpacs,showkeys,a4paper]{revtex4}
\usepackage{amsmath}
\usepackage{slashed,braket}
\usepackage{bm}
\usepackage{graphicx}

\begin{document}
\title{Meson Decays in an Extended Nambu--Jona-Lasinio model with
  Heavy Quark Flavors} 

\author{Hong-Bo Deng}

\author{Xiao-Lin Chen} 

\author{Wei-Zhen Deng}\email{dwz@pku.edu.cn}

\affiliation{School of physics and State Key Laboratory of Nuclear
  Physics and Technology, Peking University, Beijing, 100871, China}

\begin{abstract}
  In a previous work, we proposed an extended Nambu--Jona-Lasinio
  (NJL) model including heavy quark flavors. In this work, we will
  calculate strong and radiative decays of vector mesons in this
  extended NJL model, including light $\rho$, $\omega$, $K^*$,
  $\phi$ and heavy $D^*$, $D^*_s$, $B^*$, $B^*_s$. 
\end{abstract}



\maketitle

\section{Introduction}

The Nambu--Jona-Lasinio (NJL) model \cite{Nambu:1961tp, Nambu:1961fr},
in its original form as a pre-QCD theory, was constructed of nucleons
that interact via an effective two-body contact interaction. Later the
model was reinterpreted as a theory of quark degrees of freedom
\cite{Eguchi:1976iz, Kikkawa:1976fe}. The most important feature of
NJL model is the chiral symmetry of Lagrangian plus a chiral symmetry
breaking ground state. The model was generalized to $SU(3)_f$ case of
light quark flavors in refs.~\cite{Klimt:1989pm, Vogl:1989ea,
  Vogl:1991qt, Klevansky:1992qe, Ebert:1994mf}.

On the other side, for heavy quark flavors the chiral symmetry no
longer holds. However, new important symmetries such as
the spin symmetry were discovered in heavy ($Q\bar q$)-mesons
\cite{Isgur:1989vq}, which is a consequence of the order $1/m_Q$ of
spin-spin interaction in the effective quark potential
\cite{Caswell:1985ui}. In ref.~\cite{Ebert:1994tv}, The NJL model was
generalized to include heavy flavors. Both the chiral symmetry in
light meson sector and the spin symmetry in heavy meson sector were
reproduced with the vector-current interaction. The bosonization
technique was used there to obtain an effective Lagrangian of meson
degrees of freedom.

However as already shown in ref.~\cite{Klimt:1989pm}, vector-current
interaction itself is not enough to reproduce the experimental masses of
light vector mesons such as $\rho$, $K^*$ etc. Other chiral
symmetrical interactions such as the axial-vector-current one, are
needed to get satisfactory results for light meson sector. But these
additional interactions do not obey the spin symmetry in heavy meson
sector since they will generate the incorrect spin-spin interaction
that is not $1/m_Q$ suppressed. In the above work~\cite{Ebert:1994tv},
the authors just introduced two coupling constants $G_1$ and $G_2$ for
the ligth meson sector and another different coupling $G_3$ for the
heavy meson sector.

In our previous work \cite{Guo:2012tm}, we proposed a solution to
extend the NJL model to comprise the heavy quark flavors. The NJL
interactions were expanded with respect to $1/m_f$ of constituent quark
mass $m_f$ just like the expansion in the heavy quark effective theory
(HQET). Naturally the vector-current interaction is dominant while
other interactions such as the typical axial-vector-current one should
be $1/m_f$ suppressed. We had performed numerical calculations for both
the light and heavy meson sectors.  The mass spectra fit the
experimental data quite well.  The decay constants of heavy mesons were
smaller than experimental values roughly by a factor of $2$.

The strong and radiative decays provide us important information about
hadron structure.  Experimentally, the decay widths of light vector
mesons have been well measured \cite{Akhmetshin:2003zn,
  Akhmetshin:2006bx, Akhmetshin:2006sc, Fujikawa:2008ma,
  Baubillier:1984wi, Chandlee:1983hf} and so far some of decay
widths or ratios of the charmed and bottomed heavy vector mesons were
reported \cite{Abachi:1988fw, Gronberg:1995qp, Abreu:1995ky}.

Generally speaking, it is a rigid test for any model to fit the
experimental values of decay width or ratio. The most popular model
for strong decay is the ${}^3P_0$ model \cite{L1, L2}. This model has
been applied to a great number of decay processes \cite{L3, L4, L19,
  L20}. The radiative decays, mainly M1 transition which takes place
when one of the constituent quark changes its spin and radiates one
photon, has been studied in potential quark models~\cite{Jena:2010zza,
  Goity:2000dk} or from flavor symmetry~\cite{Sucipto:1987qj}. For
decays of heavy mesons, abundant works have been done in the
frameworks of chiral quark model~\cite{Goity:2000dk, Deandrea:1998hi},
potential model ~\cite{Ebert:2002xz, Colangelo:1994jc}, bag
model~\cite{Orsland:1998de}, chiral perturbation
model~\cite{Amundson:1992yp}, and QCD sum rules~\cite{Aliev:1995wi,
  Dosch:1995kw}. The decays were also studied in NJL model
\cite{Bernard:1993wf, Polleri:1996fc} and from lattice QCD
\cite{Frison:2010ws, Fu:2012tj, Becirevic:2009xp}.

In this work, we will calculate strong and radiative decays of vector
mesons in the extended NJL model with heavy flavors, including light
mesons $\rho$, $\omega$, $K^*$, $\phi$ and heavy ones $D^*$, $D^*_s$,
$B^*$, $B^*_s$.

\section{Model and Formalism}

In ref.~\cite{Guo:2012tm}, the Nambu-Jona-Lasinio model was
generalized to deal with heavy quarks as well as light ones. The
Lagrangian reads
\begin{equation}
  \mathcal{L} = \bar{\psi}(i\slashed{\partial} - \hat{m}_0) \psi 
  + \mathcal{L}_4 ,
\end{equation}
where 
\begin{equation}
  \label{four}
  \mathcal{L}_4 = G_V (\bar{\psi}\lambda_c^a \gamma_\mu \psi)^2 
  + \frac{h}{m_q m_{q^\prime}}
  [(\bar{\psi} \lambda_c^a \gamma_\mu \psi)^2
  + (\bar{\psi} \lambda_c^a \gamma_\mu \gamma_5 \psi)^2] ,
\end{equation}
describes the four-point quark-quark interaction compatible with QCD
chiral symmetry. $G_V$, of dimension (mass)$^{-2}$, and the
dimensionless $h$ were parameters fixed in the spectral
calculation. The second term on the right side in Eq.~(\ref{four})
appears as higher order correction expanded with respect to
the constituent quark mass $m_q$ similiar to the HQET expansion.  We
can rewrite Eq.~(\ref{four}) in a Fierz invariant form. For the light
sector, one has
\begin{align}
  \mathcal{L}^q_4 &= \frac{4}{9}G_V [(\bar{q}\lambda^i_f q)^2 
  + (\bar{q}i \gamma_5 \lambda^i_f q)^2] \notag \\
  &- \frac{2}{9}(G_V + \frac{h}{m_q m_{q^\prime}})
  [(\bar{q}\lambda^i_f \gamma_\mu q)^2 
  + (\bar{q} \lambda^i_f \gamma_\mu \gamma_5 q)^2].
\end{align}
where $\lambda_f^i$'s are the $U_f(3)$ generators, with $\lambda_f^0 =
\sqrt{\frac{2}{3}}I$ (where $I$ is the $3 \times 3$ unit matrix) and
the rest are Gell-Mann matrices in flavour space.  For the heavy
sector, one has
\begin{align}
  \mathcal{L}^Q_4 &= \frac{8}{9}G_V [(\bar{Q} q)^2
  + (\bar{Q}i \gamma_5 \lambda^i_f q)(\bar{q}i \gamma_5 \lambda^i_f Q)] \notag \\
  &- \frac{4}{9}(G_V + \frac{h}{m_q m_Q}) [(\bar{Q} \gamma_\mu
  q)(\bar{q} \gamma^\mu Q) + (\bar{q} \gamma_\mu \gamma_5 q)(\bar{q}
  \gamma^\mu \gamma_5 q)] ,
\end{align}
where still we have
\begin{equation}
  Tr \lambda_i \lambda_j = 2\delta_{ij}.
\end{equation}
One can see that actually we only consider the higher order $1/m_qm_Q$
suppressed interaction in vector and axial-vector channels and so the
important chiral symmetry breaking vaccum (the ground state) is
unchanged.

Using Bethe-Salpeter equation (BSE), we obtained the meson masses via
the corresponding T-matrix where the mesons appear as the poles of the
T-matrix. The meson-quark coupling constants were also obtained by further
expanding the T-matrix around the meson poles.

In this work, we will use the effective meson Lagrangian to calculate
strong and radiative decays of vector mesons. The effective
meson-quark coupling constants will be directly taken from our
previous work.  In the cases of pseudo-scalar meson and vector meson,
the corresponding effective quark couplings read
\begin{align}
  \L_{\pi q} =& -g_{\pi q} \bar\psi i\gamma_5 \bm{\tau} \psi \cdot \bm{\pi}
  -\frac{f_{\pi q}}{m_\pi} \bar\psi \gamma_\mu\gamma_5 \bm{\tau} \psi 
  \cdot \partial^\mu \bm{\pi} , \\ 
  \L_{\rho q} =& -g_{\rho q} \bar\psi \gamma_\mu \bm{\tau} \psi \cdot \bm{\rho}^\mu.
\end{align}

For the decay of a vector meson (V) into two pseudo-scalars (P), one has
\begin{equation}
  \Gamma(V \to PP) = \frac{1}{2m_V} 
  \int d\phi^{(2)} | \mathcal{M}(V \to PP) |^2,
\end{equation}
where $\int d\phi^{(2)} = \int \frac{d^3 k_1}{(2\pi)^3 2E_{k_1}}
\frac{d^3 k_2}{(2\pi)^3 2E_{k_2}} (2\pi)^4 \delta^4 (q - k_1 - k_2)$
is the standard two-body phase-space-measure. In the rest frame of the
decaying meson, the decay amplitude of the vector meson can be write
as
\begin{equation}
  \mathcal{M}(V\to PP) = \epsilon^\mu T_\mu = - \bm{\epsilon}\cdot\bm{T} ,
\end{equation}
where $\epsilon^\mu$ is the polarized vector of $V$ meson.
Then we have
\begin{equation}
  \Gamma(V \to PP) = \frac{k_c}{24\pi m_V^2} |\bm{ T}|^2 .
\end{equation}
\begin{figure}
  \begin{center}
    \includegraphics[scale= 0.5]{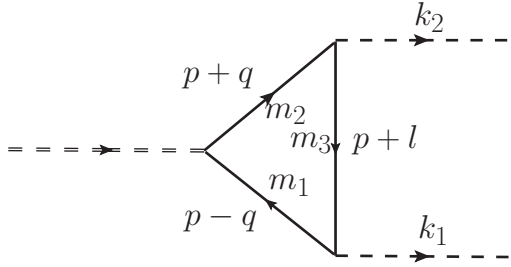}
  \end{center}
  \caption{The Feynman diagram corresponds to the strong decay
    process.}\label{strong}
\end{figure}

The strong decay process of a vector meson is shown in Feynmann
diagram Fig.~\ref{strong}, where $q = \frac{k_1 + k_2}{2} =
(\frac{m_V}{2}, \bm{0})$, $l = \frac{k_1 - k_2}{2} = (\frac{k_1^0 -
  k_2^0}{2}, \bm{k}_c)$, and $m_1$, $m_2$, $m_3$ denote the
constituent masses of the constituting quarks. Using the Feynman
rules, one can write down the expression for the decay amplitude
directly. One finds
\begin{align}
  i T^\mu =& - Tr \int\frac{d^4p}{(2\pi)^4} i g_v
  \gamma^\mu \lambda^V \frac{i}{\slashed{p} - \slashed{q} - m_1}
  i (g_1 + \frac{\tilde{g}_1}{m_1+m_3} \slashed{k}_1) 
  i \gamma_5 \lambda^{P_1} \notag\\
  & \times \frac{i}{\slashed{p} + \slashed{l} - m_3} i (g_2
  +\frac{\tilde{g}_2}{m_2 + m_3} \slashed{k}_2) i \gamma_5 \lambda^{P_2}
  \frac{i}{\slashed{p}+ \slashed{q} - m_2} .
\end{align}

For the reaction of a vector meson decays into a pseudo-scalar and a
photon ($\gamma$), $V\to P\gamma$, the decay width can be
expressed as
\begin{equation}\label{radiative}
  \Gamma(V \to P\gamma) = 
  \frac{1}{2m_V} \int d\phi^{(2)} | \mathcal{M} |^ 2,
\end{equation}
where the decay amplitude should take the form
\begin{equation}
  i \mathcal{M}(V \to P\gamma) = e 
  \epsilon^\mu(V) \epsilon^{*\nu}(\gamma) T_{\mu \nu} .
\end{equation}

\begin{figure}
\begin{center}
\includegraphics[scale= 0.5]{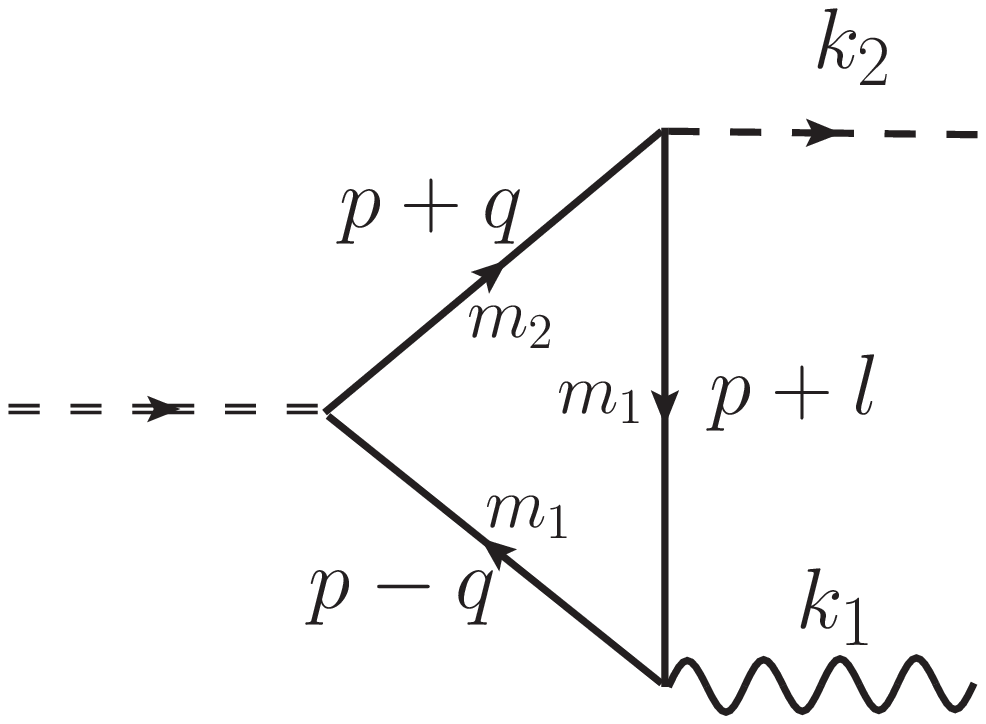}
\includegraphics[scale= 0.5]{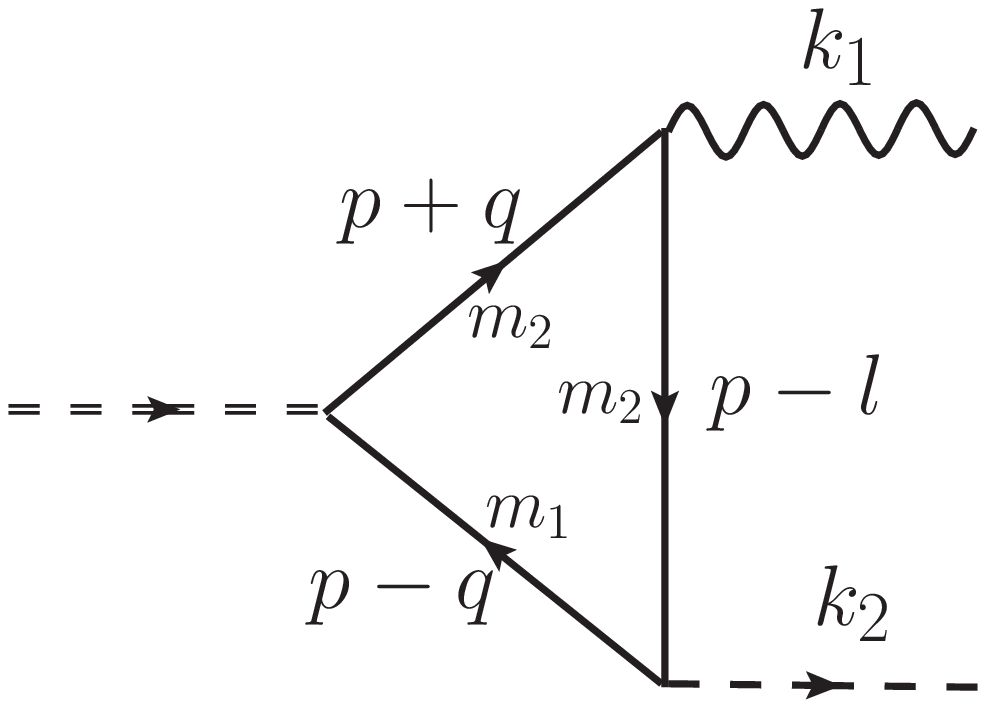}
\end{center}
\caption{The Feynman diagram corresponding to the radiative decay
  processes.}\label{radi}
\end{figure}

The Feynman diagrams of radiative decay are shown in Fig.\ref{radi}. 
We can write down the radiative decay amplitude
\begin{align*}
  T^{\mu \nu} =& Tr \int \frac{d^4p }{(2\pi)^4} i g_V \gamma^\mu
  \lambda^V \frac{i}{\slashed{p} - \slashed{q} - m_1} i \widehat{Q}
  \gamma^\nu
  \frac{i}{\slashed{p} + \slashed{l} - m_1}  \\
  & \times i (g_P + \tilde{g}_P \frac{k\!\!\!/_2}{m_1 + m_2})
  i\gamma_5 \lambda^P
  \frac{i}{\slashed{p} + \slashed{q} - m_2} \\
  &+ Tr \int \frac{d^4p }{(2\pi)^4} i g_V \gamma^\mu \lambda^V
  \frac{i}{\slashed{p} - \slashed{q} - m_1}
  i (g_P + \tilde{g}_P \frac{k\!\!\!/_2}{m_1 + m_2}) i\gamma_5 \lambda^P \\
  & \times \frac{i}{\slashed{p} - \slashed{l} - m_2} i \widehat{Q}
  \gamma^\nu \frac{i}{\slashed{p} + \slashed{q} - m_2} .
\end{align*}
In the rest frame of decaying meson, we only need the space components
of the tensor $T^{ij}$ and it can be written as
\begin{equation}
  T^{ij} = \epsilon^{ijl}  T^l_{VP\gamma}.
\end{equation}
Then we have
\begin{equation}
  \Gamma(V\to P\gamma) = \frac{\alpha k_c}{3 m_V^2} |T_{VP\gamma}|^2,
\end{equation}
where $\alpha \simeq 1/137$ is the electromagnetic fine structure
constant.

To calculate the loop integrals, we apply the three-momentum cut-off
regularization scheme to the integrals. First, we define some
useful quantities
\begin{align*}
E_p (m) =& \sqrt{\bm{p}^2 + m^2} ,\\
E_k (m) =& \sqrt{(\bm{p} + \bm{k}_c)^2 + m^2} ,\\
\omega_{1,2} =& + q^0 \pm E_p (m_1) ,\\
\omega_{3,4} =& - q^0 \pm E_p (m_2) ,\\
\omega_{5,6} =& - l^0 \pm E_k (m_3) .
\end{align*}
The $\omega_i$'s emerge as poles when the integral with respect to
$p^0$ is performed. After we integrate out $p^0$, the amplitudes can
always be represented as spatial integrals
\begin{align*}
  T =& \int^\Lambda 
  \frac{d^3 \bm{p}}{(2\pi)^3} \sum_i^{2,4,6}
  \frac{N|_{p_0 = \omega_i}}{\prod_{j \neq i} (\omega_i - \omega_j)} \\
  =& \frac{1}{4\pi^2} \int_0^\Lambda p^2dp \int_{-1}^1 dt
  \sum_i^{2,4,6} \frac{N|_{p_0 = \omega_i}}{\prod_{j \neq i} (\omega_i - \omega_j)},
\end{align*}
where $N$ represents the numerator of integrand. The 2-dimensional
integral will be performed numerically by Monte Carlo integration
method using the \texttt{vegas} routine from \texttt{gsl} library.

\section{Numerical Results}

In the previous work \cite{Guo:2012tm}, we had calculated the
pseudo-scalar and vector mesons, light and heavy, consistently in an
extended NJL model with interaction given by eq.~(\ref{four}).  The
input parameters were the current masses of light quarks and the
constituent masses of heavy quarks, the two coupling constants and the
3-dimensional cutoff.  Numerically, the parameters were set to
\begin{equation}
  \label{parameters}
  \begin{aligned}
    &m_{u/d}^0 = 2.79 \text{ MeV}, &&m_s^0 = 72.0 \text{ MeV}, \\
    &m_c = 1.62 \text{ GeV}, &&m_b = 4.94 \text{ Gev}, \\
    &\Lambda = 0.8 \text{ GeV} , &&G_V = 2.41, &&h = 0.65.
  \end{aligned}
\end{equation}
Using above parameters we obtained the constituent masses
of light quarks
\begin{align}
  &m_u = m_d = 392 \text{ MeV}, &&m_s = 542 \text{ MeV}.
\end{align}
The obtained meson-quark coupling constants, which we need
to calculate the strong and radiative decays, are given in Table
\ref{coupling}. We will use the experimental meson masses given by
Particle Date Group \cite{Nakamura:2010zzi}.

\begin{table}
  \caption{Meson-quark coupling constants.}
  \label{coupling}
  \begin{ruledtabular}
    \begin{tabular}{l cc cc cc cc cc cc c r}
      & $g_\pi$ & & $g_K$ & & $g_D$ & & $g_{D_s}$ & & $g_{B}$ & & $g_{B_s}$ & \\
      & 4.25 & & 4.32 & & 4.71 & & 5.03 & & 5.92 & & 6.69 & \\
      & $\tilde{g}_\pi$ & & $\tilde{g}_K$ & & $\tilde{g}_D$ & &
      $\tilde{g}_{D_s}$ & & $\tilde{g}_{B}$ & & $\tilde{g}_{B_s}$ & \\
      & 1.56 & & 1.61 & & 2.04 & & 2.09 & & 2.84 & & 3.11 & \\
      & $g_{\rho/\omega}$ & & $g_\phi$ & & $g_{K^*}$ & & $g_{D^*}$ & 
      & $g_{D_s^*}$ & & $g_{B^*}$ & & $g_{B_s^*}$ & \\
      & 1.29 & & 1.38 & & 1.31 & & 1.64 & & 1.83 & & 2.51 & 
      & 2.89 \\
    \end{tabular}
  \end{ruledtabular}
\end{table}

In Table \ref{light}, we show the results for the strong and radiative
decays of light vector mesons. As we can see, our results are in
qualitative agreement with the empirical values.  
\begin{table}
  \caption{Strong and radiative decay widths for light vector mesons.}
  \label{light}
  \begin{ruledtabular}
    \begin{tabular}{l c c c c}
      Decay modes & & This work & Bernard & Empirical \\
      & & & \cite{Bernard:1993wf} & \cite{Nakamura:2010zzi} \\
      \hline
      $\rho \to \pi\pi$ & MeV & 68.5 & 52.0 & 149.1 $\pm$ 0.8 \\
      $\rho^\pm \to \pi^\pm \gamma$ & keV & 21.9 & 60.1 & $68\pm 7$ \\
      $\rho^0 \to \pi^0 \gamma$ & keV & 43.9 & $-$ & $89 \pm 12$ \\
      $\omega \to \pi\gamma$ & keV & 866 & 762 & $764\pm 51$ \\
      $\phi \to K^+ K^-$ & MeV & 1.28 & $-$ & 2.08 \\
      $\phi \to K_L^0 K_S^0$ & MeV & 0.86 & $-$ & 1.46 \\
      $K^{*\pm} \to (K\pi)^\pm$ & MeV & 20.9 & 57.3 & $50.7\pm 0.9$\\
      $K^{*\pm} \to K^\pm\gamma$ & keV & 13.5 & 92.0 & $50\pm 0.5$ \\
      $K^{*0} \to K^0\gamma$ & keV & 31.3 & $ - $ & $117 \pm 10$ \\
    \end{tabular}
  \end{ruledtabular}
\end{table}
Nevertheless, quantitatively our results are systematically smaller
than the empirical values by a factor of $2$ or $3$. The discrepancy
always occurs in the NJL calculation as the model lacks the quark
confinement mechanism.  In the potential model \cite{Godfrey:1985xj},
generally the masses of light vector mesons $\rho$ or $K^*$ lay above
the constituent quark mass thresholds and still they are bound states
due to the linear confinement potential. In our calculation, the
constituent masses of light quarks are intentionally tune larger so
that the mesons are still bound states under the constituent quark
mass thresholds, even without the confinment. In another NJL
calculation \cite{Bernard:1993wf}, the smaller constituent quark
masses were used and the $\rho$ and $K^*$ vector meson was found as
the resonant poles. Then they suggested to account for the discrepancy
by introducing a renormalization factor of roughly $2$ into the light
vector meson field after have taken the higher order meson loops into
consideration.  In comparison, the numerical results from
ref.~\cite{Bernard:1993wf} are also listed in Table \ref{light}.  As
we know, the amplitudes of triangle Feynman Diagrams heavily depend on
the quarks masses when the meson masses are close to the mass
threshold. Our numerical study shows that to fit the experimental
decay width of $\rho$ demands that $2m_u$ should be very close to
$m_\rho$ and then the numerical result turns to be unstable.  We guess
that the confinement mechanism is important here for the light vector
mesons as it is critical to their formation.

Table \ref{heavy} shows the strong and radiative decay widths of heavy
vector mesons. Table \ref{branching} exhibits the branching ratios
for charmed vector mesons. It can be seen that our results agree with
the experimental values. As the empirical data are not complete, here
we also list some of other model calculation and lattice calculation
in the table for comparison. 
\begin{table}
  \caption{Strong and radiative decay widths for heavy vector mesons 
    (all in unit keV).}
  \label{heavy}
  \begin{ruledtabular}
    \begin{tabular}{l c c c c c}
      Decay Modes & This work & Kamal & Goity & Empirical \\
      & & \cite{Kamal:1992uv} & \cite{Goity:2000dk} & 
      \cite{Anastassov:2001cw, Abachi:1988fw, Gronberg:1995qp} \\
      \hline
      $D^{*\pm} \to D^\pm\pi0$ & 39.7 & 25.9 & 28.8 & \\
      $D^{*\pm} \to D^0\pi^\pm$ & 84.4 & 58.8 & 64.6 & \\
      $D^{*\pm} \to D^\pm\gamma$ & 0.7 & 1.7 & 1.4 & \\
      $D^{*\pm} \to$ all & 124.4 & 86.4 & 94.9 & 96$\pm$22 \\
      $D^{*0} \to D^0\pi^0$ & 46.5 & 42.4 & 41.6 &  \\
      $D^{*0} \to D^0\gamma$ & 19.4 & 21.8 & 32.0 & \\
      $D^{*0} \to$ all & 65.9 & 64.2 & 73.6 & $<2.1$ MeV \\
      $D^{*}_s \to D_s\gamma$ & 0.09 & 0.21 & 0.32 & $< 1.9$ MeV\\
      $B^{*\pm} \to B^\pm\gamma$ & 0.25 & $-$ & 0.74 & \\
      $B^{*0} \to B^\pm\gamma$ & 0.22 & $-$ & 0.23 & \\
      $B^{*}_s \to B_s\gamma$ & 0.10 & $-$ & 0.14 & \\
    \end{tabular}
  \end{ruledtabular}
\end{table}
\begin{table}
  \caption{Branching ratios for charmed vector mesons.}\label{branching}
  \begin{ruledtabular}
    \begin{tabular}{l c c c c c}
      Decay Modes & This work & Kamal & Goity & Empirical \\
      & & \cite{Kamal:1992uv} & \cite{Goity:2000dk} & \cite{Nakamura:2010zzi}\\
      \hline
      $D^{*\pm} \to D^\pm\pi^0$ & 31.8 & 30.0 & 30.3 & $30.7\pm 0.5$ \\
      $D^{*\pm} \to D^0\pi^\pm$ & 67.7 & 68.0 & 68.1 & $67.7\pm 0.5$ \\
      $D^{*\pm} \to D^\pm\gamma$ & 0.5 & 2.0 & 1.5 & $1.6 \pm 0.5 $ \\
      $D^{*0} \to D^0\pi^0$ & 70.6 & 66.0 & 56.5 & $61 \pm 2.9$ \\
      $D^{*0} \to D^0\gamma$ & 29.4 & 34.0 & 43.5 & $38.1 \pm 2.9$ \\
    \end{tabular}
  \end{ruledtabular}
\end{table}
In Table \ref{heavy}, our decay width of $D^{*+}$ is a little lager
than the empirical one. Numerically this can be corrected by changing
$m_c$ slightly, about 5 MeV larger. In Table \ref{branching}, our
resulted branching ratios also are in agreement with the experimental
data.  Here the numerical results are less sensitive to constituent
quark masses than that of the light meson sector. We may expect that
the calculation of strong and radiative decays for heavy mesons are
more reliable as it is well known that for heavy mesons the
confinement is less important than the one gluon exchange coulomb
potential.

\section{Summary}

We have used the extended NJL model with heavy flavors
\cite{Guo:2012tm} to calculate strong and radiative decays of vector
mesons.  It should be noted that no extra assumption and free
parameter was introduced into our present calculation.  A reasonable
agreement to the experimental data is obtained. The results of light
vector mesons may indicate that a more complex quark structure should
be considered for vector meson due to the confinement which is lacked
in NJL model.

\begin{acknowledgments}
  We would like to thank professor Shi-Lin Zhu for useful discussions.
\end{acknowledgments}

\bibliography{triangle}

\end{document}